\documentclass[twoside,twocolumn,9pt]{article}

\usepackage{extsizes}
\usepackage[super,sort&compress,comma]{natbib} 
\usepackage[version=3]{mhchem}
\usepackage[left=1.5cm, right=1.5cm, top=1.785cm, bottom=2.0cm]{geometry}
\usepackage{balance}
\usepackage{mathptmx}
\usepackage{sectsty}
\usepackage{graphicx} 
\usepackage{lastpage}
\usepackage[format=plain,justification=justified,singlelinecheck=false,font={stretch=1.125,small,sf},labelfont=bf,labelsep=space]{caption}
\usepackage{float}
\usepackage{fancyhdr}
\usepackage{fnpos}
\usepackage[english]{babel}
\addto{\captionsenglish}{%
  
}
\usepackage{array}
\usepackage{droidsans}
\usepackage{charter}
\usepackage[T1]{fontenc}
\usepackage[usenames,dvipsnames]{xcolor}
\usepackage{setspace}
\usepackage[compact]{titlesec}
\usepackage{hyperref}
\usepackage{epstopdf}
\usepackage{amssymb}

\definecolor{cream}{RGB}{222,217,201}

\begin{document}

\pagestyle{fancy}
\thispagestyle{plain}
\fancypagestyle{plain}{
\renewcommand{\headrulewidth}{0pt}
}

\makeFNbottom
\makeatletter
\renewcommand\LARGE{\@setfontsize\LARGE{15pt}{17}}
\renewcommand\Large{\@setfontsize\Large{12pt}{14}}
\renewcommand\large{\@setfontsize\large{10pt}{12}}
\renewcommand\footnotesize{\@setfontsize\footnotesize{7pt}{10}}
\makeatother

\renewcommand{\thefootnote}{\fnsymbol{footnote}}
\renewcommand\footnoterule{\vspace*{1pt}%
\color{cream}\hrule width 3.5in height 0.4pt \color{black}\vspace*{5pt}} 
\setcounter{secnumdepth}{5}

\makeatletter 
\renewcommand\@biblabel[1]{#1}            
\renewcommand\@makefntext[1]%
{\noindent\makebox[0pt][r]{\@thefnmark\,}#1}
\makeatother 
\renewcommand{\figurename}{\small{Fig.}~}
\sectionfont{\sffamily\Large}
\subsectionfont{\normalsize}
\subsubsectionfont{\bf}
\setstretch{1.125} 
\setlength{\skip\footins}{0.8cm}
\setlength{\footnotesep}{0.25cm}
\setlength{\jot}{10pt}
\titlespacing*{\section}{0pt}{4pt}{4pt}
\titlespacing*{\subsection}{0pt}{15pt}{1pt}

\fancyfoot{}
\fancyfoot[LO,RE]{\vspace{-7.1pt}\includegraphics[height=9pt]{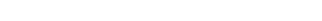}}
\fancyfoot[CO]{\vspace{-7.1pt}\hspace{11.9cm}\includegraphics{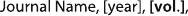}}
\fancyfoot[CE]{\vspace{-7.2pt}\hspace{-13.2cm}\includegraphics{head_foot/RF}}
\fancyfoot[RO]{\footnotesize{\sffamily{1--\pageref{LastPage} ~\textbar  \hspace{2pt}\thepage}}}
\fancyfoot[LE]{\footnotesize{\sffamily{\thepage~\textbar\hspace{4.65cm} 1--\pageref{LastPage}}}}
\fancyhead{}
\renewcommand{\headrulewidth}{0pt} 
\renewcommand{\footrulewidth}{0pt}
\setlength{\arrayrulewidth}{1pt}
\setlength{\columnsep}{6.5mm}
\setlength\bibsep{1pt}

\makeatletter 
\newlength{\figrulesep} 
\setlength{\figrulesep}{0.5\textfloatsep} 

\newcommand{\topfigrule}{\vspace*{-1pt}%
\noindent{\color{cream}\rule[-\figrulesep]{\columnwidth}{1.5pt}} }

\newcommand{\botfigrule}{\vspace*{-2pt}%
\noindent{\color{cream}\rule[\figrulesep]{\columnwidth}{1.5pt}} }

\newcommand{\dblfigrule}{\vspace*{-1pt}%
\noindent{\color{cream}\rule[-\figrulesep]{\textwidth}{1.5pt}} }

\makeatother

\twocolumn[
  \begin{@twocolumnfalse}
{\includegraphics[height=30pt]{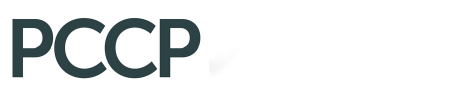}\hfill\raisebox{0pt}[0pt][0pt]{\includegraphics[height=55pt]{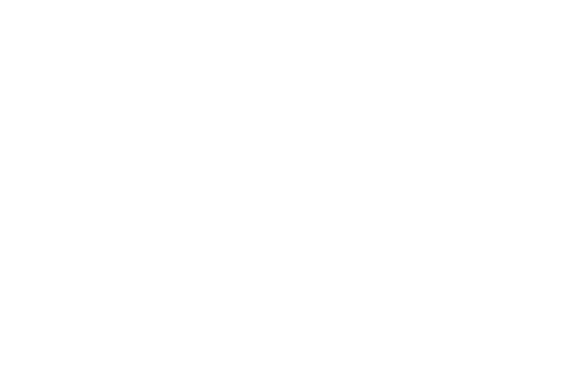}}\\[1ex]
\includegraphics[width=18.5cm]{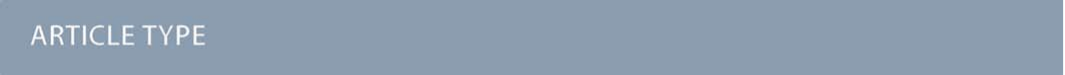}}\par
\vspace{1em}
\sffamily
\begin{tabular}{m{4.5cm} p{13.5cm} }

\includegraphics{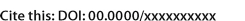} & \noindent\LARGE{\textbf{Symmetry and Topology in Wavepacket Dynamics near Conical Intersections}} \\
\vspace{0.3cm} & \vspace{0.3cm} \\

 & \noindent\large{Christopher Daggett\textit{$^{a}$}, Jonathan Rawlinson\textit{$^{b}$}, Lukas Muechler$^{\ast}$\textit{$^{a}$}} \\

\includegraphics{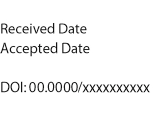} & \noindent\normalsize{Whether topology directly shapes chemical dynamics remains an open question in theoretical chemistry. The issue arises because degeneracies of adiabatic electronic states can generate nontrivial topological structure, and such degeneracies are common in polyatomic molecules. Existing work has largely emphasized static characterizations and dynamical studies of low-energy, highly symmetric models. Here we develop a symmetry-based analysis of nonadiabatic dynamics in two-state conical-intersection models that is predictive without invoking topological invariants. We show that the nodal-line structures associated with dynamics near a conical intersection are robust in highly symmetric settings, but should not in general be expected to persist once the relevant symmetry is broken.} \\

\end{tabular}

 \end{@twocolumnfalse} \vspace{0.6cm}
]

\renewcommand*\rmdefault{bch}\normalfont\upshape
\rmfamily
\section*{}
\vspace{-1cm}

\footnotetext{\textit{$^{a}$Department of Chemistry, Penn State University, State College, USA. E-mail: cdd5353@psu.edu, lfm5572@psu.edu}}
\footnotetext{\textit{$^{a}$Department of Physical Sciences, Nottingham Trent University, Nottingham, UK. E-mail: Jonathan.rawlinson@ntu.ac.uk}}


\section{Introduction}

Conical intersections (CIs) are central features in the study of nonadiabatic molecular dynamics. In photochemical reactions they act as efficient pathways for nonradiative decay of electrons from excited states and are common occurrences on the potential energy surfaces (PESs) of polyatomic molecules~\cite{yarkony1998conical,bernardi1996potential,domcke2012role,worth2004beyond}. 
Their importance has motivated the development of a broad range of theoretical and computational methods.~\cite{domcke2012role,bestnacpractice,ferretti1996quantum}
A common challenge shared by these methods is the need to describe nuclear motion in regions where the electronic basis changes rapidly and where the Born-Oppenheimer approximation breaks down.

Beyond methodological challenges, CIs also raise conceptual questions about the possible role of geometric and topological effects in nuclear dynamics.
In the adiabatic representation, the electronic eigenfunctions associated with a real two-state CI change sign when transported around a loop enclosing the intersection. This sign change is commonly described as a geometric- or Berry phase of $\pi$~\cite{LonguetHigginsSignFlip, BerryPhase,mead1979determination,mead1992geometric,yarkony2001nuclear}. The sign change reflects the static nontrivial topology of the corresponding real adiabatic wavefunction. 

This static property is often invoked to interpret the dynamics of wavepackets traversing a CI at low energies. ~\cite{yuan2018observation,schon1995geometric,joubert2013geometric}
A nuclear wavepacket approaching a CI splits into two components which proceed to travel on opposite sides of the intersection.  Upon passing the CI fully the wavepacket components do not recombine, instead they remain separated by a line of vanishing nuclear density. This is attributed to accumulation of differing phases by the wavepacket components which upon completely encircling the CI results in a phase difference of $\pi$. Consequently, they experience pure destructive interference along a line in nuclear configuration space. This interpretation has become a standard way of rationalizing nodal structures in minimal two-dimensional conical-intersection models.\cite{mead1992geometric,yarkony2001nuclear,ryabinkin2017geometric} \par

The purpose of this work is to examine and clarify what part of this interpretation is topological and what part is dynamical. We argue that the topology of the adiabatic state and the appearance of nodal lines in time-dependent wavepacket dynamics are distinct. The former concerns the global structure of a single real adiabatic state. The latter concerns the symmetry and interference properties of the full vibronic wavefunction. In particular, as we discuss here, for a highly symmetric model the nodal line can be derived directly from conservation of a discrete vibronic symmetry under unitary time evolution.\par

In Sec.~\ref{s2a} and ~\ref{s2b} we review the static topology of the canonical real two-state CI model and clarify the relation of this model to two topological invariants, the quantized Berry phase and the first Stiefel-Whitney class. We continue in Sec.~\ref{s2c} by illustrating the role of off-diagonal nonadiabatic couplings (NACs) in the calculation of the Berry phase and reviewing the structure and implications of these objects. Sec.~\ref{s2d} contrasts the full two-state electronic space and the single adiabatic state subspace. Sec.~\ref{s2e} concludes by formulating the symmetry principle responsible for nodal-line formation in the full vibronic wavefunction. In Sec.~\ref{s3} we provide computational details used to perform nonadiabatic dynamics simulations in this work. Sec.~\ref{s4} provides the results of wavepacket dynamics simulations for two model systems. ~Sec.~\ref{s5} provides an interpretation of simulation results and a discussion of their implications.

\section{Background}
\label{s2}

\subsection{The Two-State Conical Intersection Model}
\label{s2a}

The essential properties of an isolated CI may be captured by a minimal two-state linear crossing model. For the sake of simplicity we choose to represent this model as a real, two-state, diabatic electronic Hamiltonian.\par

\begin{equation}
\label{ci_ham}
    \textbf{H}_{elec}(R_1,R_2)= 
    \begin{bmatrix}
    \epsilon(R_1,R_2)+\alpha R_1 & \beta R_2 \\
    \beta R_2 &  \epsilon(R_1,R_2)-\alpha R_1 \\
    \end{bmatrix}
\end{equation}

We take the $R_1,R_2$ to be the branching plane coordinates. The adiabatic energies of $\textbf{H}_{elec}(R_1,R_2)$ likewise take on a simple form.

\begin{equation}
\label{adi_ener}
    E_\pm(R_1,R_2) = \epsilon(R_1,R_1)\pm\sqrt{\alpha^2 R_2^2+\beta^2R_2^2}
\end{equation}

For the remainder of the present discussion we omit the $\epsilon(R_1,R_2)$ terms as they change only the topography of the PES and have no impact on topological properties of interest. The adiabatic states corresponding to $\textbf{H}_{elec}(R_1,R_2)$ are degenerate only at the origin and are linearly separated through coupling terms away from this point.\par

For further investigation, we switch to the more convenient polar coordinate system.

\begin{equation}
\label{polar_coord}
\alpha R_1=R\sin\phi,
\qquad
\beta R_2=R\cos\phi ,
\end{equation}

The Hamiltonian then appears as:

\begin{equation}
\label{polar_ham}
    \textbf{H}_{elec}(\phi)= R
    \begin{bmatrix}
    \cos \phi & \sin \phi \\
    \sin \phi &  -\cos \phi \\
    \end{bmatrix}
\end{equation}

Setting aside the issue of choosing a specific phase, one valid choice of eigenvectors is:

\begin{equation}
\label{polar_evecs}
\psi_-(\phi)=
\begin{bmatrix}
-\sin(\phi/2)\\
\cos(\phi/2)
\end{bmatrix},
\qquad
\psi_+(\phi)=
\begin{bmatrix}
\cos(\phi/2)\\
\sin(\phi/2)
\end{bmatrix}.
\end{equation}

The half-angle dependence leads to:

\begin{equation}
\label{sign_flip}
\psi_{\pm}(\phi+2\pi)=-\psi_{\pm}(\phi).
\end{equation}

Mathematically, this sign-change represents the underlying topology of a real line bundle.
A real line bundle can be thought of as tracking a single real adiabatic eigenstate around a loop in nuclear coordinate space, while asking whether its sign can be chosen consistently all the way around.
For a loop enclosing the CI, this line bundle is nontrivial, i.e. no globally smooth, single-valued, real representative of the adiabatic states can be chosen. This obstruction can be removed only by giving up one of these conditions.~\cite{mead1979determination}
For example, a smooth periodic gauge can be introduced by complexifying the state,

\begin{equation}
\widetilde{\psi}_{\pm}(\phi)=e^{i\phi/2}\psi_{\pm}(\phi).
\end{equation}

This restores $2\pi$ periodicity, but the state is no longer real. Conversely, if one insists on a real representation, a single-valued choice must contain a sign discontinuity.~\cite{kendrick2003geometric}

\subsection{Berry phase and the first Stiefel-Whitney class}
\label{s2b}

The topology of the real adiabatic line bundle is diagnosed by topological invariants.
Within the single-state adiabatic description, the point of degeneracy is ill-defined. This is because the projector onto one nondegenerate adiabatic state is not defined at the CI. The local base space is therefore the punctured plane.

\begin{equation}
\mathbb{R}^2\setminus\{0\},
\end{equation}

As a result, any closed loop that encircles the CI cannot be continuously shrunk to a point without crossing the degeneracy. The real adiabatic line bundle over this loop is topologically classified by the first Stieffel-Whitney class $w_1 \in \mathbb{Z}_2$.
For a loop that encloses the CI once, this invariant is nontrivial, i.e. $w_1=1$
For a loop that does not enclose the CI or encloses an even number of equivalent sign-changing intersections, the invariant is trivial, i.e. $w_1=0$.

In the literature, this invariant is commonly expressed as a Berry phase $\gamma$.~\cite{ahn2019stiefel}  For one isolated real adiabatic state on a closed loop around a CI,

\begin{equation}
w_1=\gamma/\pi \mod 2 .
\end{equation}

The Berry phase and the first Stiefel-Whitney class therefore contain the same $\mathbb{Z}_2$ information in this setting. However, the Stiefel-Whitney formulation makes explicit that the topology belongs to the real adiabatic line bundle. The Berry phase gives an equivalent description after complexification.

In practice, calculation of the Berry phase is straightforward in the complex basis.

\begin{equation}
\gamma=\oint i\langle \widetilde{\psi}_-(\phi)|
\partial_{\phi}\widetilde{\psi}_-(\phi)\rangle\,d\phi=\pi\mod 2\pi,
\end{equation}

when the loop encircles an odd number of CIs.

\subsection{Off-diagonal nonadiabatic coupling and the finite-loop distinction}
\label{s2c}

While the complex basis is preferred for evaluation of these invariants, most quantum chemistry codes use a real basis due to computational efficiency.
For an isolated two-state system, an indirect method of computation exists using only real adiabatic states,

\begin{equation}
\gamma_{12} = \oint A_{12}\cdot d\textbf{R} = n\pi , n \in \mathbb{Z},
\end{equation}
where 
\begin{equation}
A_{12}(\textbf{R})=\langle \psi_1(\textbf{R})|\nabla_\textbf{R}\psi_2(\textbf{R})\rangle.
\end{equation}
are off-diagonal NACs.~\cite{kendrick2003geometric, yarkony1998conical}

This quantity is closely related to the geometric phase but is generally not identical to the Berry phase of a single adiabatic state.

This relation between Berry phases and $\gamma_{12}$ becomes transparent after complexifying the two real adiabatic states. 
Define
\begin{equation}
\chi_{\pm}(\textbf{R})
=
\frac{1}{\sqrt{2}} \bigl(\psi_1 (\textbf{R})\pm i\psi_2(\textbf{R}) \bigr).
\end{equation}
A direct calculation gives
\begin{equation}
i\langle \chi_{\pm}|\nabla_\textbf{R}\chi_{\pm}\rangle
=
\mp A_{12}(\textbf{R}).
\end{equation}
Thus the integral of $A_{12}$ can be interpreted as the Berry phase of these complexified states. In the ideal real two-state CI model, this integral gives the same $\mathbb{Z}_2$ information as the Berry phase of an individual adiabatic state and the first Stiefel-Whitney class.

This equivalence is special to the minimal setting.~\cite{yarkony1996consequences,kendrick2002properties,mead1982conditions} In realistic molecular systems, the full off-diagonal NAC may contain both an irrotational contribution associated with the Longuet-Higgins sign change and a solenoidal contribution that is not removable by a gauge transformation,

\begin{equation}
A_{12}=A_{12}^{\mathrm{irr}}+A_{12}^{\mathrm{sol}} .
\end{equation}

The irrotational contribution integrates to a multiple of $\pi$,

\begin{equation}
\oint A_{12}^{\mathrm{irr}}\cdot d\textbf{R}=n\pi ,
\end{equation}

whereas the solenoidal contribution is not generally quantized,

\begin{equation}
\oint A_{12}^{\mathrm{sol}}\cdot d\textbf{R}\neq n\pi .
\end{equation}

In the limit of a vanishingly small loop around an isolated conical intersection, the solenoidal contribution becomes negligible and the full $A_{12}$ integral approaches a multiple of $\pi$:

\begin{equation}
\lim_{C\to 0}\oint_C A_{12}\cdot d\textbf{R}=n\pi .
\end{equation}

For finite loops, however, deviations from quantization can occur. The Stiefel-Whitney class remains quantized, but the finite-loop integral of the full off-diagonal NAC need not be.

This distinction is important for dynamics. A wavepacket samples a finite region of the PES and is affected by the details of the potential energy landscape, not only by the infinitesimal topology near the CI. This is particularly important for ab-initio calculations, where deviations from quantization can be significant.

\subsection{From one adiabatic state to a two-state vibronic description}
\label{s2d}

The topological obstruction described above belongs to a single adiabatic state. Once both electronic states are retained, the relevant electronic space is the direct sum

\begin{equation}
E=L_-\oplus L_+ .
\end{equation}

Here $L_-$ and $L_+$ denote the one-dimensional real adiabatic subspaces associated with the lower and upper electronic states. In the minimal two-state CI model, this rank-2 real bundle is trivial. The original diabatic basis in which Eq.~\ref{ci_ham} is written provides a smooth real two-state basis everywhere, including at the CI.

The obstruction is therefore not an obstruction to choosing a smooth real basis for the full two-state electronic space. It is an obstruction to globally splitting this smooth two-state space into two smooth, single-valued, real adiabatic states. This distinction is essential for nonadiabatic dynamics. The propagated vibronic wavefunction contains nuclear amplitudes on both electronic states, and its symmetry properties are those of the full electron-nuclear state, not of one adiabatic state alone.

The conventional geometric-phase explanation of nodal-line formation can now be rephrased more carefully. Imagine a wavepacket approaching a CI with low energy such that no significant transfer to the excited adiabatic state can take place. Then the wavepacket passing through a CI will split into two branches whose relative sign is affected by the electronic sign change. Their recombination can produce a nodal line. This is a useful approach for the canonical single-state point of view relevant for energies below the CI, and has been discussed in detail elsewhere.~\cite{yuan2018observation,schon1995geometric,joubert2013geometric} However, it is not the only way to derive the nodal structure. In the symmetric models considered here, the same conclusion follows directly from unitary symmetry preservation as we will show below

\subsection{Symmetry principle for nodal-line formation}
\label{s2e}

Assume that the electronic Hamiltonian satisfies a reflection symmetry,

\begin{equation}
U H_{\mathrm{el}}(R_1,R_2)U^{-1}
=
H_{\mathrm{el}}(R_1,-R_2),
\qquad
U^2=\mathbb{1} .
\end{equation}

Let $\hat P_2$ reflect the nuclear coordinate $R_2\to -R_2$, and define the combined vibronic symmetry operator

\begin{equation}
\hat G=\hat P_2 U .
\end{equation}

If

\begin{equation}
    {[}\hat{H},\hat{G}{]}=0,
\end{equation}

then unitary time evolution preserves the total symmetry sector. Thus,

\begin{equation}
\hat G\Psi(R,0)=\eta\Psi(R,0),
\qquad
\eta=\pm1,
\end{equation}

implies

\begin{equation}
\hat G\Psi(R,t)=\eta\Psi(R,t)
\end{equation}

for all times.

Expand the vibronic wavefunction as

\begin{equation}
\Psi(R,t)
=
\chi_g(R,t)\phi_g(R)
+
\chi_e(R,t)\phi_e(R).
\end{equation}

On the symmetry line $R_2=0$, the electronic states can be chosen as eigenstates of $U$:

\begin{equation}
U\phi_n(R_1,0)
=
\lambda_n(R_1)\phi_n(R_1,0),
\qquad
\lambda_n(R_1)=\pm1 .
\end{equation}

The conserved total symmetry then requires

\begin{equation}
\chi_n(R_1,-R_2,t)
=
\eta\,\lambda_n(R_1)\,
\chi_n(R_1,R_2,t).
\end{equation}

Therefore the nuclear component in channel $n$ is even in $R_2$ when $\eta\lambda_n(R_1)=+1$, and odd when $\eta\lambda_n(R_1)=-1$. In the odd case,

\begin{equation}
\chi_n(R_1,0,t)=0 ,
\end{equation}
so a nodal line is enforced by symmetry.
For the canonical CI Hamiltonian,
\begin{equation}
H_{\mathrm{el}}(R_1,R_2)
=
R_1\sigma_z+R_2\sigma_x,
\qquad
U=\sigma_z ,
\end{equation}
the symmetry line is $R_2=0$, where
\begin{equation}
H_{\mathrm{el}}(R_1,0)=R_1\sigma_z .
\end{equation}
The lower-state electronic symmetry changes across the CI:
\begin{equation}
\lambda_g(R_1<0)=+1,
\qquad
\lambda_g(R_1>0)=-1 .
\end{equation}
If a Gaussian wavepacket is initialized on the lower state for $R_1<0$, is even in $R_2$, and has momentum along the symmetry line, then the total vibronic state has a definite symmetry. After passage through the CI, the lower electronic state has the opposite symmetry eigenvalue. To preserve the total symmetry, the lower-state nuclear component must become odd in $R_2$, forcing a nodal line along $R_2=0$.

This argument extends immediately to excited-state dynamics. If population transfers into a channel whose electronic symmetry eigenvalue changes between the initial and final regions, the nuclear component must change parity and develop a node. If the electronic symmetry eigenvalue does not change, no symmetry-enforced node is required. The nodal-line criterion is therefore controlled by the change in electronic symmetry character, not by whether the channel is labeled ground or excited.

The symmetry argument makes a concrete prediction for perturbed models. Close to the CI, an approximately symmetric Hamiltonian can still generate a nodal structure because the local dynamics resembles the symmetric model. Away from the CI, symmetry-breaking terms in the PES can change the relative amplitudes, phases, and trajectories of the split wavepacket components. The nodal line may therefore appear transiently near the intersection but need not persist at later times.

This provides the central distinction tested below. Static topology fixes the sign structure of an adiabatic state around the CI. Symmetry conservation fixes the parity of nuclear components in a full vibronic wavefunction. The persistence of a nodal line in finite-time dynamics depends on how the PES guides and recombines the wavepacket branches. The simulations below examine this distinction by comparing symmetry-preserving and symmetry-breaking dynamics.

\section{Computational Details}
\label{s3}

A series of simulations were performed to provide a foundation for further discussion of the potential impact of topological properties on nuclear wavepacket dynamics. We describe here the explicit models employed and highlight some of the key features of the associated PESs and wavefunctions. \par

The full nuclear-electronic Hamiltonian is a sum of the nuclear kinetic energy term and a diabatic electronic Hamiltonian.

\begin{equation}
    \textbf{H}_{full}(\textbf{R}) = -\frac{1}{2}\nabla_\textbf{R}^2 +\textbf{H}_{elec}(\textbf{R}) 
\end{equation}

\begin{figure}[th]
    \centering
    \includegraphics[width=\columnwidth]{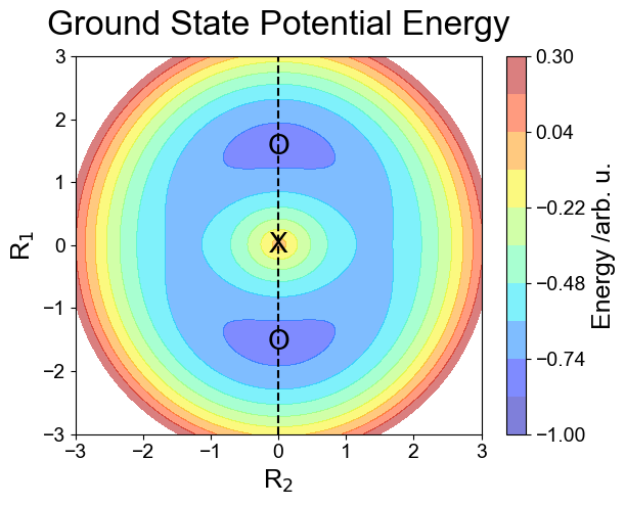}
    \caption{The ground state potential energy surface for the symmetric Hamiltonian $\textbf{H}_{symm}$. The dotted line along the $R_1$ axis denotes the mirror symmetry. The "X" marks the location of the conical intersection, and the "O"s mark the equivalent global minima.} 
    \label{fig:pes_grd}
\end{figure}

The coordinates $\textbf{R}=(R_1,R_2)$ upon which the Hamiltonian depends are taken to be the two nuclear degrees of freedom in the branching plane of a CI. The diabatic electronic Hamiltonians used were all of the general form:

\begin{equation}
    \textbf{H}_{elec}(\textbf{R}) = \textbf{H}_{symm}(\textbf{R})+\textbf{H}_{per}
\end{equation}
In the above expression $\textbf{H}_{symm}$ is a two-state linear crossing model of the type in Eq.~\ref{ci_ham}.
\begin{equation}
    \textbf{H}_{symm}= 
    \begin{bmatrix}
    a(R_1^2+R_2^2)+bR_1 & ic R_2 \\
    -icR_2 &  a(R_1^2+R_2^2)-bR_1 \\
    \end{bmatrix}
\end{equation}
The terms quadratic in $R_1,R_2$ along the diagonal are included in order to act as a confining potential which bounds the region within which dynamics take place. The parameters $a=0.32$, $b=1.0$, $c=0.9$ were chosen to produce a potential energy surface with two equivalent minima along the $R_1$ axis, connected by a pair of transition states along the $R_2$ axis. Here, we choose a complex basis for computational reasons. The results of the dynamics do not depend on the choice of gauge for the electronic states. An illustrative example of the ground state adiabatic PES corresponding to $\textbf{H}_{symm}$ is given in Fig.~\ref{fig:pes_grd}.

The term $\textbf{H}_{per}$ is a constant matrix which is included for symmetry-breaking. Explicitly,

\begin{equation}
    \textbf{H}_{per}= 
    \begin{bmatrix}
    0 & i d \\
    -i d  &  0 \\
    \end{bmatrix}
\end{equation}
where $d = 0.15$.


The total vibronic wavefunction is expressed as an expansion into the electronic adiabatic basis $\phi_i$ which is complete for our model systems.
\begin{equation}
\label{eq:wf_full}
    \Psi(t,\textbf{R}) = \sum_i\chi_i(\textbf{R},t) \phi_i(\textbf{R}) 
\end{equation}
The nuclear wavefunction at initialization ($t=0$) is chosen to be a weighted gaussian wavepacket.
\begin{equation}
\chi_i(\textbf{R};\boldsymbol\alpha,\textbf{R}_0,\textbf{p}_0) = \frac{c_i}{2\pi\sigma}e^{i\textbf{p}_0\textbf{R}-\boldsymbol\alpha(\textbf{R}-\textbf{R}_0)^2}
\end{equation}
The initialization parameters are the width parameter ($\boldsymbol\alpha$), initial momentum ($\textbf{p}_0$), and initial position ($\textbf{R}_0$). The weighting coefficients $c_i$ are chosen to produce the desired initial adiabatic state populations. For simulations in this work initialization is either purely in the ground state ($c_1=1)$ or purely in the excited state ($c_2=1)$.
The nuclear wavefunctions are complex. We initialize the real part as a radially symmetric gaussian, while the imaginary part stems from the momentum. If the initial momentum is along the line of reflection, the wavefunction is symmetric.
The method of time propagation for the dynamics is second order split-operator Fourier transform (SOFT).~\cite{kosloff1988time}

A summary of the wavepacket dynamics as we performed them will now be given, with detailed examples in the following section. A spatial grid representing $\textbf{R}$ was initialized with $256\times256$ points corresponding to the region $R_1,R_2\in {[}-3,3{]}$. This resolution was found to be sufficiently accurate for the analysis in this paper. The electronic eigenvectors are found for every gridpoint via pointwise numerical diagonalization of $\textbf{H}_{elec}$, and from these the corresponding diabatic states are obtained. We then define the nuclear wavepacket at each gridpoint and construct the full wavefunction according to Eq.~\ref{eq:wf_full}. The parameters $\textbf{p}_0,\textbf{R}_0$ are used to center the wavepacket as desired, while $\boldsymbol\alpha$ dictates its spatial extent and $\textbf{c}=(c_1,c_2)$ sets the intial state populations. Time propagation is then performed in the diabatic basis using the second order split-operator Fourier transform method. The selected timestep was $\Delta t=0.1$ for all simulations.

\section{Phenomenology}
\label{s4}

Below we present the results of a series of simulations where a nuclear wavepacket encounters a CI of the PES in the course of dynamical evolution. The key motivation behind the selected model, parameters, and initial conditions was to investigate the nodal line phenomenon and its connection to the symmetry of the vibronic wavefunction. To this end we present three fundamentally different scenarios: Fully symmetric ground state dynamics, fully symmetric excited state dynamics, and ground state dynamics on a symmetry broken PES.\par

In order to have a quantitative measure of the proposed conserved total symmetry of the wavepacket we introduce the symmetry correlation function $\xi(t)$ as a surrogate for the symmetry properties of the nuclear wavefunction.

\begin{equation}
\label{symm_corr}
    \xi_i(t) = n\int_\textbf{S}\chi _i^*(R_1,R_2,t)\chi _i(R_1,-R_2,t)d\textbf{S}
\end{equation}

In the above expression $\textbf{S}$ is defined as a half plane:

\begin{equation}
    \textbf{S}=\{R_1,R_2|R_2<0\}
\end{equation}

The coefficient $n$ normalizes $\chi_i(\textbf{R})$ over $\textbf{S}$. Practically Eq.~\ref{symm_corr} is evaluated numerically. This amounts to calculating the nuclear wavefunction symmetry eigenvalues with respect to reflection over the $R_1$ axis at each gridpoint in the half-plane $\textbf{S}$, followed by summation. Defined in this way, $\xi_i(t)$ may be interpreted as a measure for the action of the symmetry operator on the $i-th$ nuclear wavefunction. 

\subsection{Symmetric Case - Ground State}
\label{s4a}

\begin{figure*}[th]
    \centering
    \includegraphics[width=\textwidth]{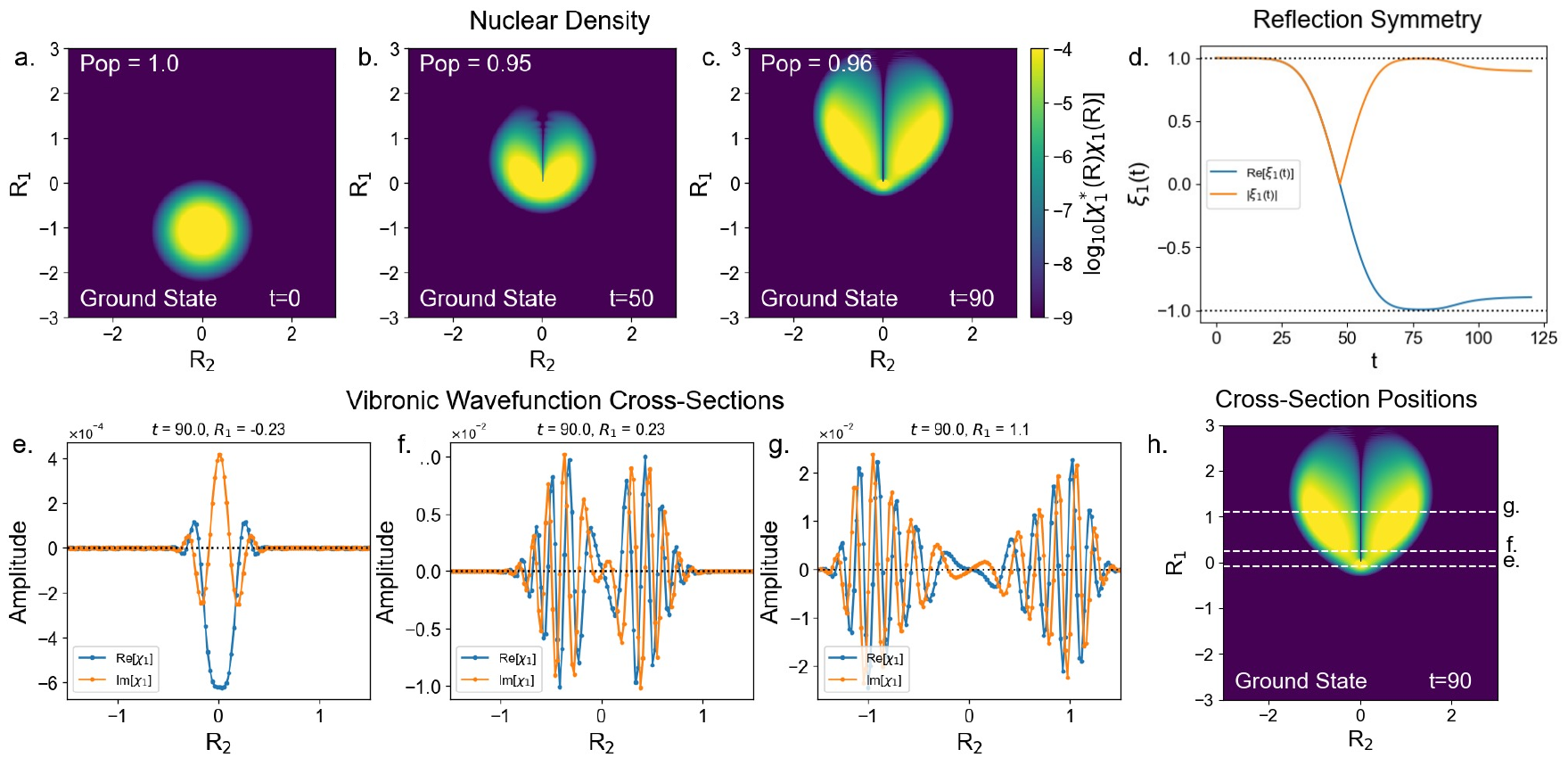}
    \caption{(a-c) Snapshots of ground state nuclear density during wavepacket dynamics on a symmetric PES. (a) The initial radially symmetric wavepacket in one ground state minima. (b) Near the CI the partial bifurcation of the wavepacket is apparent. (c) Almost complete bifurcation of the wavepacket is evident, with a clearly identifiable nodal line along the positive $R_1$ axis. (d) Symmetry surrogate (Eq.~\ref{symm_corr}) of the nuclear wavepacket with respect to reflection over the $R_1$ axis showing a smooth transition from even to odd symmetry.(e-g) Cross-sections of the nuclear wavepacket at the final timestep. e) The cross-section shows even symmetry in the $R_1<0$ half-plane. (f,g) The cross-sections in the $R_1<0$ half-plane show odd symmetry. (h) Depiction of cross-section positions.}
    \label{fig:symm_grd_revised}
\end{figure*}

Our first simulation serves to create a baseline for the dynamics of a nuclear wavepacket, initialized on the adiabatic ground state, which encounters a CI in the course of dynamics. For this purpose the most highly symmetric diabatic Hamiltonian is selected to generate the PES.

\begin{equation}
    \textbf{H}_{elec}(\textbf{R}) = \textbf{H}_{symm}(\textbf{R})
\end{equation}

The wavepacket was initialized centered about the local minima of the ground state PES at $\textbf{R}_0={[}-1.5,0{]}$ with a momentum of $\textbf{p}_0={[}55,0{]}$ and an $\boldsymbol\alpha={[}4,4{]}$. $\textbf{R}_{0}$ was chosen such that the nuclear wavepacket is spatially symmetric with respect to the reflection over the $R_1$ axis and $\textbf{p}_0$ is chosen such that the wavepacket may pass through the region of the PES containing the CI with minimal nonadiabatic transfer to the adiabatic excited state. Snapshots of the nuclear density at several time points in the simulation are shown in Fig~\ref{fig:symm_grd_revised}\par

Starting from $t=0$ the nuclear wavepacket moves in the positive $R_1$ direction. As it encounters the CI an area of zero nuclear density forms along the $R_1$ axis, effectively bifurcating the wavepacket. There is also a minimal transfer of nuclear density into the excited state which at no point exceeds $9\%$. 

Finally, after the bifurcated ground-state wavepacket exits the CI region, the nodal-line structure remains well preserved. At longer times, the excited-state nuclear density also relaxes to the ground state.

Analysis of the nuclear wavefunction makes explicit how electronic-state symmetry constrains the nodal-line structure. Fig.~\ref{fig:symm_grd_revised}d displays $\xi_1$ over the course of the simulation, in which there is an overall change in the nuclear wavepacket symmetry from even to odd. From approximately $t=20$ to $t=75$ $\xi_1$ changes smoothly in value from 1 to -1. Comparing this to plots of the nuclear density in Fig~\ref{fig:symm_grd_revised}a-c shows a correlation between this change in symmetry and the formation of the nodal line. In Fig~\ref{fig:symm_grd_revised}a the nuclear wavepacket is even and shows no nodal line at initialization. At the final time, shown in Fig~\ref{fig:symm_grd_revised}a the nuclear wavepacket has odd symmetry and shows a clear nodal line. At an intermediate time (Fig~\ref{fig:symm_grd_revised}b the nuclear density is partially even and partially odd with $\xi_1$ not quantized due to the presence of nuclear density for $R_1 < 0$ originating from an even nuclear wavefunction.\par
To further illustrate the conservation of symmetry Fig.~\ref{fig:symm_grd_revised}e-h displays cross sections of the  vibronic wavefunction with respect to $R_1$ for fixed values of $R_2$ at the final time step. Fig.~\ref{fig:symm_grd_revised}e shows that for $R_1<0$ the wavefunction remains even. Nuclear density in this region originates partially from decay of the excited state branch of the wavepacket. For $R_1>0$ the wavefunction becomes odd as shown in Fig.~\ref{fig:symm_grd_revised}f,g. 
These results correspond to a maintained symmetry of the full vibronic wavefunction.

\subsection{Symmetric Case - Excited State}
\label{s4b}

\begin{figure*}[th]
    \centering
    \includegraphics[width=\textwidth]{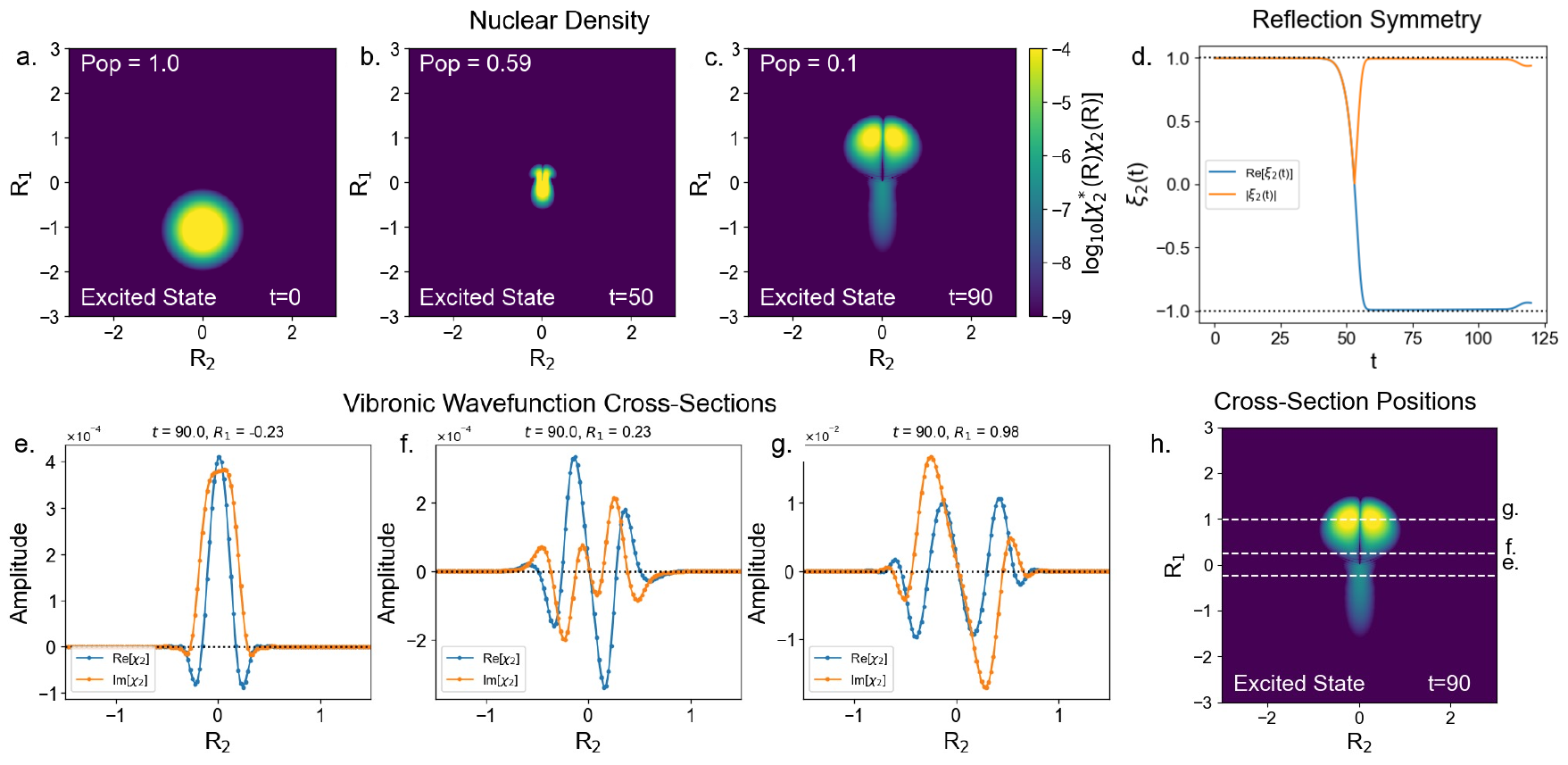}
    \caption{(a-c) Snapshots of excited state nuclear density during wavepacket dynamics on a symmetric PES. a) The initial radially symmetric wavepacket in one ground state minima. b) Near the CI significant population transfer to the ground state occurs, along with partial bifurcation of the excited state branch. c) At the final time there exists a nodal line along the positive $R_1$ axis. d) Symmetry surrogate (Eq.~\ref{symm_corr}) of the excited state nuclear wavepacket with respect to reflection over the $R_1$ axis.(e-g) Cross-sections of the nuclear wavepacket at the final timestep. e) The cross-section shows even symmetry in the $R_1<0$ half-plane. f,g) The cross-sections in the $R_1<0$ half-plane show odd symmetry. h) Depiction of cross-section positions.}
    \label{fig:symm_exc_revised}
\end{figure*}

To illustrate the applicability of symmetry analysis, we now initialize a wavepacket in the excited state and study its relaxation dynamics through the CI under the lens of symmetry. In contrast to the Berry-phase arguments discussed earlier, the symmetry analysis trivially generalizes to dynamics involving multiple states and remains predictive.

\begin{figure*}[th]
    \centering
    \includegraphics[width=\textwidth]{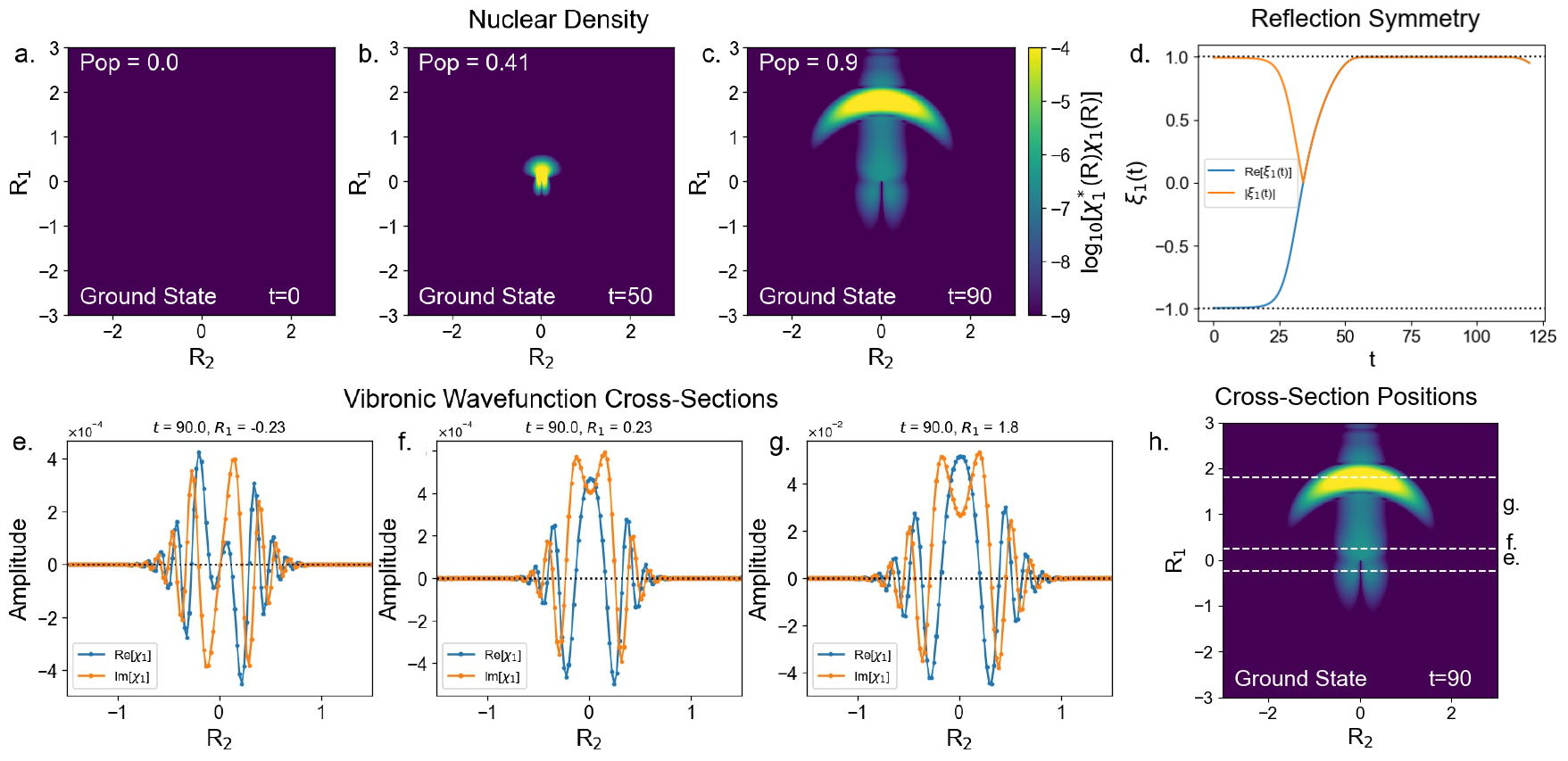}
    \caption{(a-c) Snapshots of ground state nuclear density during wavepacket dynamics on a symmetric PES. (a) The initial radially symmetric wavepacket near a ground state minima. (b) As the nuclear density passes the CI a bifurcation forms and deviation from symmetry with respect to the $R_1$ axis is evident. (c) At the final time the wavepacket is nearly bifurcated along a curve in the $R_1>0$ half-plane. d) Symmetry surrogate (Eq.~\ref{symm_corr}) of the ground state nuclear wavepacket with respect to reflection over the $R_1$ axis smoothly declines, showing no definite symmetry after initialization. (e-g) None of the cross-sections display a definite even or odd symmetry. (h) Depiction of cross-section positions. }
    \label{fig:symm_grd_revisedB}
\end{figure*}

For this purpose the same diabatic Hamiltonian was used as in the first simulation, $\textbf{H}_{symm}$. The wavepacket was initialized centered about the nuclear coordinate corresponding to the local minima of the ground state PES, designated in the previous section as $\textbf{R}_0={[}-1.5,0{]}$. A momentum of $\textbf{p}_0={[}0,0{]}$ was chosen as the excited state gradient alone is sufficient to ensure direct and efficient passage through the CI. $\boldsymbol\alpha={[}4,4{]}$. Snapshots of the nuclear density at several time points in the simulation are shown for the excited state in Fig.~\ref{fig:symm_exc_revised}a-c and for the ground state in Fig.~\ref{fig:symm_grd_revisedB}a-c.\par

Starting from $t=0$ the nuclear wavepacket accelerates in the positive $R_1$ direction towards the CI, which is located at the global minima of the excited state PES. Upon reaching the CI the  majority of nuclear density transitions to the lower adiabatic state and proceeds with no nodal line formation for $R_1>0$. The total percent transmission to the ground state at the final time step is $90\%$. The branch on the excited state moves past the CI and shows a clear nodal line along the $R_1$ axis for $R_1>0$.\par  


Analysis of the excited state nuclear density shows agreement with the prediction of conserved total symmetry. Referring to Fig.~\ref{fig:symm_exc_revised}d, a transition from even to odd symmetry of the excited state wavepacket is apparent as $\xi_2$ decays smoothly from $\xi_2=1$ at $t=45$ to $\xi_2=-1$ at $t=60$. The nuclear density plot in  Fig~\ref{fig:symm_exc_revised}a,c demonstrate this behavior visually, showing that the nuclear wavepacket is initially even and spatially localized within the $R_1<0$ half-plane while at the final time it is odd and localized in the $R_1>0$ half-plane. At an intermediate time (Fig~\ref{fig:symm_exc_revised}b) the nuclear wavepacket is distributed between the two half-planes and has a non-quantized symmetry eigenvalue $\xi_2\neq\pm1$ overall, but shows distinctive symmetry character within each half-plane.\par
Cross-sections of the excited state nuclear wavefunction in Fig.~\ref{fig:symm_exc_revised}e-g demonstrate the symmetry basis for nodal line formation. Fig.~\ref{fig:symm_exc_revised}e demonstrates the even symmetry in the $R_1<0$ half-plane, while f.,g. display the odd symmetry for the $R_1>0$ half-plane. This behavior is unchanged by the significant transfer of nuclear density to the ground state in the course of dynamics.\par
For the ground state nuclear density, as shown in Fig.~\ref{fig:symm_grd_revisedB}d, the branch of the original wavepacket which transitions to the ground state exhibits no nodal line in the $R_1>0$ half-plane and retains even symmetry in this region.However, in Fig~\ref{fig:symm_grd_revisedB}b a nodal line can be identified in the $R_1<0$ half-plane. The cross-section of nuclear density in the ground state Fig.~\ref{fig:symm_grd_revisedB}e-g, shows behavior consistent with the symmetry predictions analagous to the excited state.\par

Overall the excited-state simulation demonstrates that the formation of a nodal line can be predicted solely from the symmetry of the electronic eigenstates and the initial symmetry of the nuclear wavepacket, even if multiple electronic states are involved.

\subsection{Asymmetric Case - Ground State}
\label{s4c}

\begin{figure}[h!]
    \centering
    \includegraphics[width=\columnwidth]{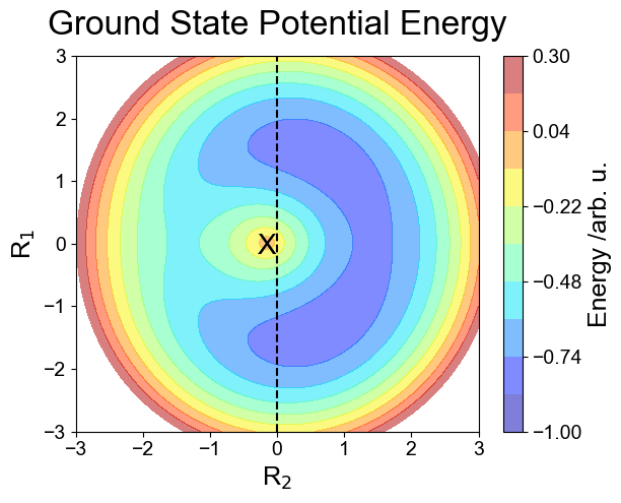}
    \caption{The ground state potential energy surface of $\textbf{H}_{symm}+\textbf{H}_{per}$ used for the third dynamics simulation. The dotted line along the $R_1$ axis visibly no longer corresponds to a  reflection symmetry. The "X" marks the location of the conical intersection}
    \label{fig:pes_asymm_grdA}
\end{figure}

The final simulation was designed to assess the behavior of the nodal line phenomena when the symmetry of the Hamiltonian with respect to reflection over the $R_1$ axis is broken. To break the symmetry a term $\textbf{H}_{per}$ is added to the the symmetric Hamiltonian $\textbf{H}_{symm}$. This addition has the effect of shifting the position of the CI slightly along the $R_2$ axis, but more importantly it shifts the diabatic coupling so that its magnitude is not equal on either side of the $R_1$ axis. This results in two distinct paths around the CI, with different energetic barrier heights as shown in Fig. ~\ref{fig:pes_asymm_grdA}.

The wavepacket was initialized centered about the $R_1$ axis of the ground state PES at $\textbf{R}=-1.5$ with a momentum of $\textbf{p}_0=55$. These initial conditions were chosen such that the initial wavepacket is positioned with its center in line with the CI, and so that it has sufficient momentum to ensure both non-equivalent paths about the CI are energetically accessible. Snapshots of the nuclear density at several time points in the simulation are shown in Fig~\ref{fig:asymm_grd_revisedB}. \par

Starting from $t=0$ the wavepacket proceeds directly towards the CI parallel to the $R_1$ axis. Upon reaching the CI, the wavepacket bifurcates into unequal parts. The smaller part corresponds to that portion of the incident wavepacket which must cross the larger potential energy barrier. A nodal line structure forms as in the other simulations, however it is visibly not as well maintained over time compared to the previous simulations. \par 

Fig.~\ref{fig:asymm_grd_revisedB}d shows the loss of reflection symmetry as the initial value of $\xi_1=1$ begins to decrease almost immediately, and remains non-quantized for all subsequent time steps. Referring to ~\ref{fig:asymm_grd_revisedB}a-c shows the progress of the nuclear wavepacket. At $t=0$ the wavepacket is even by design. At the final time there is a region of very low nuclear density which nearly bifurcates the wavepacket, however it is along a curve in coordinate space rather than a line, and does not appear as well maintained as in the previous two simulations. At an intermediate time (Fig.~\ref{fig:asymm_grd_revisedB}b) a partial bifurcation is present, and the nodal line is distinctly identifiable.\par

The cross sections of the nuclear wavepacket in Fig.~\ref{fig:asymm_grd_revisedB}e-g show the divergence from symmetry. Near the CI, the $R_2$ cross section in Fig.~\ref{fig:asymm_grd_revisedB}g shows that the wavefunction no longer preserves clear antisymmetry about the expected nodal line.. This lack of antisymmetry is even more pronounced in the final cross section (Fig.~\ref{fig:asymm_grd_revisedB}d).

\begin{figure*}[th]
    \centering
    \label{fig:asymm_grd_revisedB}
    \includegraphics[width=\textwidth]{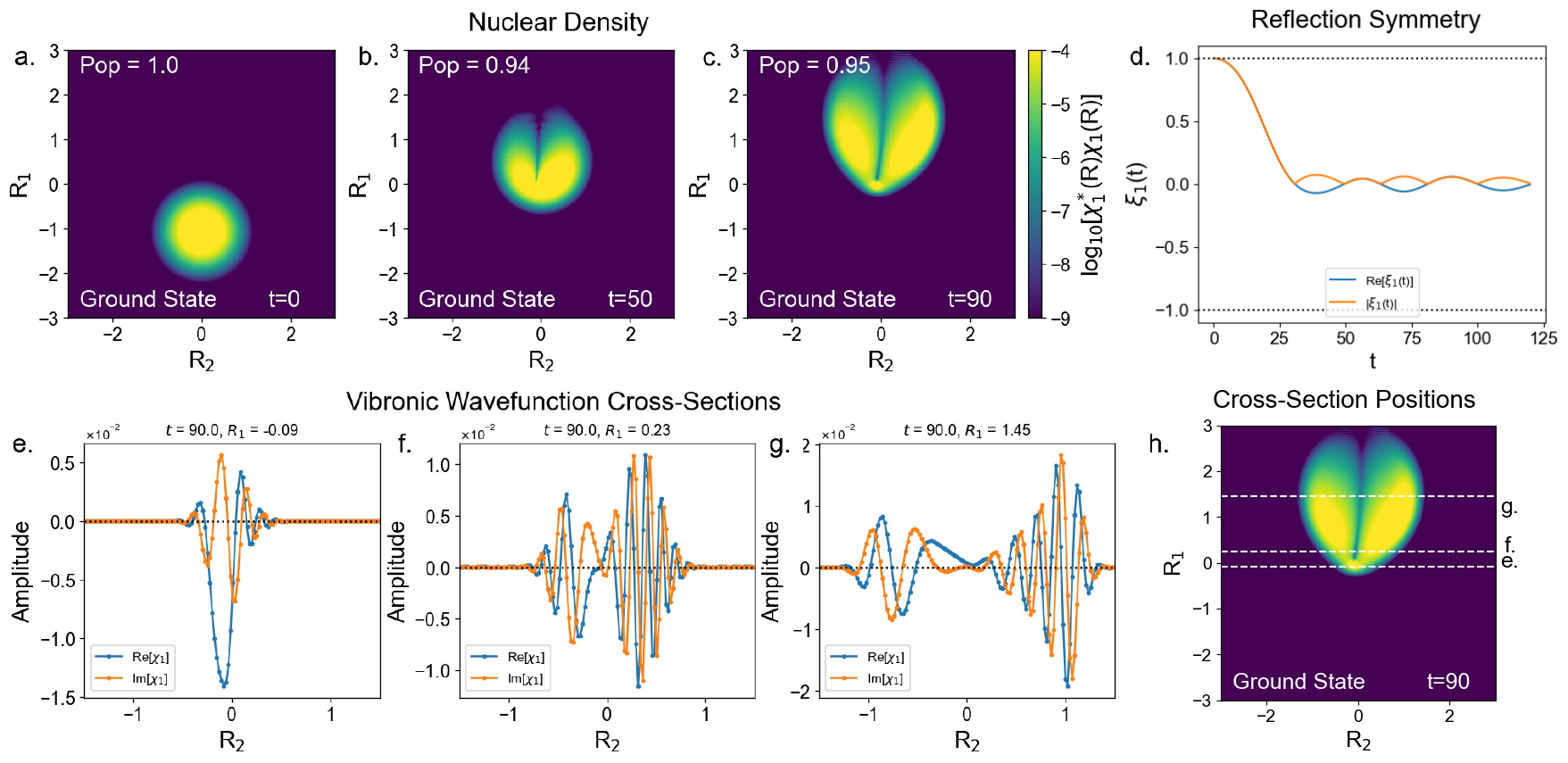}
    \caption{(a-c) Snapshots of ground state nuclear density during wavepacket dynamics on a symmetry broken PES. (a) The wavepacket is initialized entirely in the excited state. (b) Nuclear density from the excited state spreads on the ground state surface. The beginning of a bifurcation along the negative $R_1$ axis is apparent. (c) At the final time no nodal line is observable along the positive $R_1$ axis, while a bifurcation can still be identified along the negative $R_1$ axis. (d) Symmetry surrogate (Eq.~\ref{symm_corr}) of the nuclear wavepacket with respect to reflection over the $R_1$ axis smoothly transitions from odd to even. (e) The cross-section shows odd symmetry in the $R_1<0$ half-plane. (f,g) The cross-sections in the $R_1<0$ half-plane show even symmetry. (h) Depiction of cross-section positions. }
    \label{fig:asymm_grd_revisedB}
\end{figure*}

The loss of symmetry in the PES results in a loss of symmetry in its gradient thus subjecting the wavepacket to non-symmetric forces. From ~\ref{fig:asymm_grd_revisedB}d it can be observed that $\xi_1$ begins to decrease from the very first time step, which is before the wavepacket reaches the CI. This is in contrast to the previous simulations, where $\xi_1$ only decreases once the wavepacket reaches the CI, and where this behavior results from differing, but individually well-maintained,  symmetry in the $R_1<0$ and $R_1>0$ half-planes. Without a symmetry in the electronic Hamiltonian, unitary time evolution does not conserve the initial symmetry of the wavepacket, and thus does not result in a maintained nodal line. 

To further quantify the symmetry breaking, we write the nuclear wavefunction for the ground state in terms of even and odd components with respect to the line of symmetry of the symmetric Hamiltonian

\begin{equation}
\begin{split}
        \chi_{even} (\textbf{R}) & = \frac{1}{2} \left[\chi_0(R_1,R_2)+ \chi_0(R_1,-R_2) \right] \\
        \chi_{odd} (\textbf{R}) & = \frac{1}{2}  \left[ \chi_0(R_1,R_2) - \chi_0(R_1,-R_2) \right]
\end{split}
\end{equation}

For the times shown in Fig.~\ref{fig:asymm_grd_revisedB}b\&c, we find that the wavefunction is nearly evenly split between even and odd sectors, with about $50\%$ of the norm in the odd sector and $50\%$ in the even sector at $t = 50$. Since the odd sector carries the exact reflection node whereas the even sector fills it, the trench observed in the density is naturally understood as the remnant of a large odd-sector contribution rather than an exact nodal line of the full wavefunction. 

\section{Conclusions}
\label{s5}

In this work we propose a symmetry approach for predicting the outcome of wavepacket dynamics in two-level single CI systems. In such a system if the Hamiltonian is invariant under a combined operation consisting of a nuclear reflection and an electronic symmetry operation, then a wavepacket initialized in a definite symmetry sector remains in that sector. Along the symmetry line passing through the CI, the electronic symmetry character of an adiabatic state changes across the intersection. The nuclear component in that channel must then change parity to preserve the total vibronic symmetry. This parity change forces a nodal line. This argument does not require identifying the nodal line as a consequence of a topological invariant. Rather, it follows from symmetry preservation of the full electron-nuclear wavefunction. \par

This distinction is important for two reasons. First, once nonadiabatic dynamics are considered, the wavefunction generally contains amplitude on more than one electronic state. The relevant object is then a multicomponent vibronic wavefunction rather than a single adiabatic electronic state. Second, wavepacket dynamics samples finite regions of the PES. Quantized topological statements about infinitesimal loops around a CI do not, by themselves, determine the persistence of nodal structures at later times. The latter depends on symmetry, wavepacket splitting, coherence, and the potential-driven motion of the nuclear components. \par

The results of a series of dynamics simulations demonstrate the use of symmetry to predict nodal line formation. The first simulation (Sec.\ref{s4a}) reproduces the well-known nodal line phenomena for dynamics that occur primarily on the ground adiabatic state. Analysis demonstrates that as a wavepacket traveling on a single state passes the CI the formation of a nodal line is predictable from the change in the symmetry of the associated electronic state. The second simulation (Sec.\ref{s4b} expands this idea by presenting dynamics that begin on the excited state, and end with branches of nuclear density in both excited and ground state. Additionally, at the final timestep for both excited and ground state there is a nonzero nuclear density present in regions of the coordinate space corresponding to different electronic symmetry. Analysis confirms that the static symmetry of the electronic states is sufficient to predict the behavior of all branches of the nuclear wavepacket. This leads us to propose that in highly symmetric systems no appeal to topological properties of the system is necessary, as the symmetry of the system is entirely predictive of the behavior of the nuclear wavepacket. \par

Two-state single CI systems host nontrivial topology in the form of a nontrivial first Stiefel-Whitney class or equivalently a quantized Berry phase. In the adiabatic, ultracold regime and in highly symmetric model systems this topology has been investigated and found to be crucial to understand dynamics.
However, for reactions in the non-adiabatic regime taking place at elevated temperatures and energies above the CI, the topology of a single crossing does not need to be invoked. 
However, there is ample ground to explore the role of topology and other related concepts such as quantum geometry in dynamics. For example, Euler-Class like invariants can classify the distribution of NACs due to the presence of multiple CIs, and can be used to treat degenerate ground states.~\cite{daggett2024toward} Generalizations of these models and other topological invariants are possible, however have not been explored in a molecular setting.~\cite{ahn2019failure}. \par


\section*{Acknowledgements}
L.M. and C.D. would like to acknowledge Dr. Felipe Hernandez and Dr. Chaoxing Liu for stimulating discussions and valuable feedback in the course of this work.
L.M. and C.D. acknowledge support from U.S. Department of Energy, Office of Science, Office of Basic Energy Sciences, CPIMS program, under award number DE-SC0025352.

\section*{Author contributions}

\section*{Conflicts of interest}

There are no conflicts to declare.

\section*{Data availability}

Data for this article in the form of .npy files containing wavefunction information are publicly available through Zenodo at \url{https://doi.org/10.5281/zenodo.21084052}. Python code for generating figures from wavefunction files are publicly available through Github at: \url{https://github.com/chrisdagg0005/CI_PaperData}.

\balance

\bibliography{lit} 

@article{LonguetHigginsSignFlip,
 ISSN = {00804630},
 URL = {http://www.jstor.org/stable/78955},
 author = {H. C. Longuet-Higgins},
 journal = {Proc. R. Soc. Lon. Ser. A},
 number = {1637},
 pages = {147--156},
 publisher = {The Royal Society},
 title = {The Intersection of Potential Energy Surfaces in Polyatomic Molecules},
 urldate = {2026-05-21},
 volume = {344},
 year = {1975}
}

@article{BerryPhase,
 ISSN = {00804630},
 URL = {http://www.jstor.org/stable/2397741},
 author = {M. V. Berry},
 journal = {Proc. R. Soc. Lon. Ser. A},
 number = {1802},
 pages = {45--57},
 publisher = {Royal Society},
 title = {Quantal Phase Factors Accompanying Adiabatic Changes},
 urldate = {2026-05-21},
 volume = {392},
 year = {1984}
}

@article{yuan2018observation,
  title={Observation of the geometric phase effect in the H+ HD→ H2+ D reaction},
  author={Yuan, Daofu and Guan, Yafu and Chen, Wentao and Zhao, Hailin and Yu, Shengrui and Luo, Chang and Tan, Yuxin and Xie, Ting and Wang, Xingan and Sun, Zhigang and others},
  journal={Science},
  volume={362},
  number={6420},
  pages={1289--1293},
  year={2018},
  publisher={American Association for the Advancement of Science}
}

@article{schon1995geometric,
  title={Geometric phase effects and wave packet dynamics on intersecting potential energy surfaces},
  author={Sch{\"o}n, J{\"o}rg and K{\"o}ppel, Horst},
  journal={J. Chem. Phys.},
  volume={103},
  number={21},
  pages={9292--9303},
  year={1995},
  publisher={American Institute of Physics}
}

@article{joubert2013geometric,
  title={Geometric phase effects in low-energy dynamics near conical intersections: A study of the multidimensional linear vibronic coupling model},
  author={Joubert-Doriol, Lo{\"\i}c and Ryabinkin, Ilya G and Izmaylov, Artur F},
  journal={J. Chem. Phys.},
  volume={139},
  number={23},
  pages={234103},
  year={2013},
  publisher={AIP Publishing}
}

@article{kendrick2003geometric,
  title={Geometric phase effects in chemical reaction dynamics and molecular spectra},
  author={Kendrick, Brian K},
  journal={J. Phys. Chem. A},
  volume={107},
  number={35},
  pages={6739--6756},
  year={2003},
  publisher={ACS Publications}
}

@article{mead1992geometric,
  title={The geometric phase in molecular systems},
  author={Mead, C Alden},
  journal={Rev. Mod. Phys.},
  volume={64},
  number={1},
  pages={51},
  year={1992},
  publisher={APS}
}

@article{worth2004beyond,
  title={Beyond Born-Oppenheimer: Molecular dynamics through a conical intersection},
  author={Worth, Graham A and Cederbaum, Lorenz S},
  journal={Annu. Rev. Phys. Chem.},
  volume={55},
  number={1},
  pages={127--158},
  year={2004},
  publisher={Annual Reviews}
}

@article{mead1979determination,
  title={On the determination of Born--Oppenheimer nuclear motion wave functions including complications due to conical intersections and identical nuclei},
  author={Mead, C Alden and Truhlar, Donald G},
  journal={J. Chem. Phys.},
  volume={70},
  number={5},
  pages={2284--2296},
  year={1979},
  publisher={American Institute of Physics}
}

@article{yarkony2001nuclear,
  title={Nuclear dynamics near conical intersections in the adiabatic representation: I. The effects of local topography on interstate transitions},
  author={Yarkony, David R},
  journal={J. Chem. Phys.},
  volume={114},
  number={6},
  pages={2601--2613},
  year={2001},
  publisher={American Institute of Physics}
}

@article{ryabinkin2017geometric,
  title={Geometric phase effects in nonadiabatic dynamics near conical intersections},
  author={Ryabinkin, Ilya G and Joubert-Doriol, Lo{\"\i}c and Izmaylov, Artur F},
  journal={Acc. Chem. Res.},
  volume={50},
  number={7},
  pages={1785--1793},
  year={2017},
  publisher={ACS Publications}
}

@article{yarkony1998conical,
  title={Conical intersections: Diabolical and often misunderstood},
  author={Yarkony, David R},
  journal={Acc. Chem. Res.},
  volume={31},
  number={8},
  pages={511--518},
  year={1998},
  publisher={ACS Publications}
}

@article{domcke2012role,
  title={Role of conical intersections in molecular spectroscopy and photoinduced chemical dynamics},
  author={Domcke, Wolfgang and Yarkony, David R},
  journal={Annu. Rev. Phys. Chem.},
  volume={63},
  number={1},
  pages={325--352},
  year={2012},
  publisher={Annual Reviews}
}

@article{ferretti1996quantum,
  title={Quantum mechanical and semiclassical dynamics at a conical intersection},
  author={Ferretti, A and Granucci, Giovanni and Lami, A and Persico, Maurizio and Villani, G},
  journal={J. Chem. Phys.},
  volume={104},
  number={14},
  pages={5517--5527},
  year={1996},
  publisher={American Institute of Physics}
}

@article{ahn2019stiefel,
  title={Stiefel--Whitney classes and topological phases in band theory},
  author={Ahn, Junyeong and Park, Sungjoon and Kim, Dongwook and Kim, Youngkuk and Yang, Bohm-Jung},
  journal={Chin. Phys. B},
  volume={28},
  number={11},
  pages={117101},
  year={2019},
  publisher={Chinese Physical Society and IOP Publishing Ltd}
}

@article{bernardi1996potential,
  title={Potential energy surface crossings in organic photochemistry},
  author={Bernardi, Fernando and Olivucci, Massimo and Robb, Michael A},
  journal={Chem. Soc. Rev.},
  volume={25},
  number={5},
  pages={321--328},
  year={1996},
  publisher={The Royal Society of Chemistry}
}

@article{yarkony1996consequences,
  title={On the consequences of nonremovable derivative couplings. I. The geometric phase and quasidiabatic states: A numerical study},
  author={Yarkony, David R},
  journal={J. Chem. Phys.},
  volume={105},
  number={23},
  pages={10456--10461},
  year={1996},
  publisher={American Institute of Physics}
}

@article{kendrick2002properties,
  title={Properties of nonadiabatic couplings and the generalized Born--Oppenheimer approximation},
  author={Kendrick, Brian K and Mead, C Alden and Truhlar, Donald G},
  journal={Chem. Phys.},
  volume={277},
  number={1},
  pages={31--41},
  year={2002},
  publisher={Elsevier}
}

@article{mead1982conditions,
  title={Conditions for the definition of a strictly diabatic electronic basis for molecular systems},
  author={Mead, C Alden and Truhlar, Donald G},
  journal={J. Chem. Phys.},
  volume={77},
  number={12},
  pages={6090--6098},
  year={1982},
  publisher={American Institute of Physics}
}

@article{daggett2024toward,
  title={Toward a Topological Classification of Nonadiabaticity in Chemical Reactions},
  author={Daggett, Christopher and Yang, Kaijie and Liu, Chao-Xing and Muechler, Lukas},
  journal={Chem. Mater.},
  volume={36},
  number={8},
  pages={3479--3489},
  year={2024},
  publisher={ACS Publications}
}

@article{ahn2019failure,
  title={Failure of Nielsen-Ninomiya theorem and fragile topology in two-dimensional systems with space-time inversion symmetry: application to twisted bilayer graphene at magic angle},
  author={Ahn, Junyeong and Park, Sungjoon and Yang, Bohm-Jung},
  journal={Phys. Rev. X},
  volume={9},
  number={2},
  pages={021013},
  year={2019},
  publisher={APS}
}

@article{kosloff1988time,
  title={Time-dependent quantum-mechanical methods for molecular dynamics},
  author={Kosloff, Ronnie},
  journal={J. Phys. Chem.},
  volume={92},
  number={8},
  pages={2087--2100},
  year={1988},
  publisher={ACS Publications}
}

@article{bestnacpractice, 
    place={Boulder, CO, USA}, 
    title={Best practices for nonadiabatic molecular dynamics simulations [Article v1.0]}, 
    volume={7}, 
    url={https://livecomsjournal.org/index.php/livecoms/article/view/v7i1e4157}, 
    DOI={10.33011/livecoms.7.1.4157}, 
    number={1}, 
    journal={Living Journal of Computational Molecular Science}, 
    author={Prlj, Antonio and Taylor, Jack T. and Janoš, Jiří and Lognon, Elise and Hollas, Daniel and Slavíček, Petr and Agostini, Federica and Curchod, Basile F. E.}, year={2026},
    month={Jan.}, 
    pages={4157}}
\bibliographystyle{rsc.bst} 

\end{document}